\documentclass[aoas,MSNbibl,nameyear,dvips]{arximspdf}
\usepackage{graphicx}
\doi{10.1214/14-AOAS726} 
\volume{8}
\issue{2}
\pubyear{2014}
\firstpage{747}
\lastpage{776}

\makeatletter
\newcommand{\PP}{\mathbb{P}}
\makeatother

\begin{document}
\begin{frontmatter}

\title{Clustering South African households based on their asset status using latent variable models}
\runtitle{Clustering South African households}

\begin{aug}
\author[a]{\fnms{Damien}~\snm{McParland}\ead[label=e1]{damien.mcparland@ucd.ie}\thanksref{t1,m1}},
\author[a]{\fnms{Isobel~Claire}~\snm{Gormley}\corref{}\ead[label=e2]{claire.gormley@ucd.ie}\thanksref{t1,m1}},
\author[b]{\fnms{Tyler~H.}~\snm{McCormick}\ead[label=e4]{tylermc@u.washington.edu}\thanksref{t3,m2}},
\author[c]{\fnms{Samuel~J.}~\snm{Clark}\ead[label=e3]{samclark@u.washington.edu}\thanksref{t2,m2,m3,m4,m5}},
\author[d]{\fnms{Chodziwadziwa~Whiteson}~\snm{Kabudula}\ead[label=e5]{cho@agincourt.co.za}\thanksref{m3}}
\and
\author[d]{\fnms{Mark~A.}~\snm{Collinson}\ead[label=e6]{mark@agincourt.co.za}\thanksref{m3,m5}}
\runauthor{D. McParland et al.}
\address[a]{D. McParland\\
I.~C. Gormley\\
School of Mathematical Sciences\\
University College Dublin\\
Belfield, Dublin 4\\
Ireland\\
\printead{e1}\\
\phantom{E-mail:\ }\printead*{e2}}
\address[b]{T.~J. McCormick\\
Department of Sociology\\
University of Washington\\
Seattle, Washington 98195-3340\\
USA\\
and\\
Centre for Statistics\\
\quad and the Social Sciences\\
University of Washington\\
Seattle, Washington 98195-4320\\
USA\\
\printead{e4}\hspace*{16pt}}
\address[c]{S. J. Clark\\
Department of Sociology\\
University of Washington\\
Seattle, Washington 98195-3340\\
USA\\
and\\
MRC/Wits Rural Public Health\\
\quad and Health Transitions\\
\quad Research Unit (Agincourt)\\
School of Public Health\\
University of the Witwatersrand\\
Johannesburg\\
South Africa\\
and\\
Institute of Behavioral Science (IBS)\\
University of Colorado at Boulder\\
Boulder, Colorado 80302\\
USA\\
and\\
INDEPTH Network\\
Accra\\
Ghana\\
\printead{e3}}
\address[d]{C.~W. Kabudula\\
M.~A. Collinson\\
MRC/Wits Rural Public Health\\
\quad and Health Transitions\\
\quad Research Unit (Agincourt)\\
School of Public Health\\
University of the Witwatersrand\\
Johannesburg\\
South Africa\\
and\\
INDEPTH Network\\
Accra\\
Ghana\\
\printead{e5}\\
\phantom{E-mail:\ }\printead*{e6}}

\affiliation{University College Dublin\thanksmark{m1},
University of Washington\thanksmark{m2},\break
University of the Witwatersrand\thanksmark{m3},
University of Colorado at Boulder\thanksmark{m4}
and INDEPTH Network\thanksmark{m5}}

\thankstext{t1}{Supported by Science Foundation Ireland, Grant number
09/RFP/MTH2367.}
\thankstext{t2}{Supported by NIH Grants K01 HD057246, R01 HD054511, R24
AG032112.}
\thankstext{t3}{Supported by NIH Grant R01 HD054511 and a Google
Faculty Research Award.}

\end{aug}

\received{\smonth{8} \syear{2013}}
\revised{\smonth{1} \syear{2014}}


\begin{abstract}
The Agincourt Health and Demographic Surveillance System has since 2001
conducted a biannual household asset survey in order to quantify
household socio-economic status (SES) in a rural population living in
northeast South Africa. The survey contains binary, ordinal and nominal
items. In the absence of income or expenditure data, the SES landscape
in the study population is explored and described by clustering the
households into homogeneous groups based on their asset status.

A model-based approach to clustering the Agincourt households, based on
latent variable models, is proposed. In the case of modeling binary or
ordinal items, item response theory models are employed. For nominal
survey items, a factor analysis model, similar in nature to a
multinomial probit model, is used. Both model types have an underlying
latent variable structure---this similarity is exploited and the
models are combined to produce a hybrid model capable of handling mixed
data types. Further, a mixture of the hybrid models is considered to
provide clustering capabilities within the context of mixed binary,
ordinal and nominal response data. The proposed model is termed a
mixture of factor analyzers for mixed data (MFA-MD).

The MFA-MD model is applied to the survey data to cluster the Agincourt
households into homogeneous groups. The model is estimated within the
Bayesian paradigm, using a Markov chain Monte Carlo algorithm.
Intuitive groupings result, providing insight to the different
socio-economic strata within the Agincourt region.
\end{abstract}

%

\end{frontmatter}

\section{Introduction}

The Agincourt Health and Demographic Surveillance System (HDSS) [\citet
{kahn07}] continuously monitors the population of 21
villages located in the Bushbuckridge subdistrict of Mpumalanga
Province in
northeast South Africa. This is a rural population living in what was,
during Apartheid, a black ``homeland.'' The Agincourt HDSS was
established in the early 1990s with the purpose of guiding the
reorganization of South Africa's health system. Since then the goals of
the HDSS have evolved and now it contributes to evaluation of national
policy at population, household and individual levels. Here, the aim is
to study the socio-economic status of the households in the Agincourt region.

Asset-based wealth indices are a common way of quantifying wealth in
populations for which alternative methods are not feasible [\citet
{vyas06}], such as when income or expenditure data are unavailable.
Households in the study area have been surveyed biannually since 2001
to elicit an accounting of assets similar to that used by the
Demographic and Health Surveys [\citet{rutstein04}] to construct a wealth
index. The SES landscape is explored by analyzing the most recent
survey of assets for each household. The resulting data set contains
binary, ordinal and nominal items.

The existence of SES strata or clusters is a well established concept
within the sociology literature. \citet{weeden12}, \citet{erikson92} and
\citet{svalfors06}, for example, expound the idea of SES clusters. \citet
{alkema08} consider a latent class analysis approach to exploring SES
clusters within two of Nairobi's slum settlements; they posit the
existence of 3 and 4 poverty clusters in the two slums, respectively. In
a similar vein, here the aim is to examine the SES clustering structure
within the set of Agincourt households, based on the asset status
survey data. Interest lies in exploring the substantive differences
between the SES clusters. Thus, the scientific question of interest can
be framed as follows: what are the (dis)similar features of the SES
clusters in the set of Agincourt households? This paper aims to answer
this question by appropriately clustering the Agincourt households
based on asset survey data. The resulting socio-economic group
membership information will be used for targeted health care
projects and for further surveys of the different socio-economic
groups. The SES strata could also serve as valuable inputs to other
analyses such as mortality models, and will serve as a key tool in
studying poverty dynamics.

To uncover the clustering structure in the Agincourt region, a model is
presented here which facilitates clustering of observations in the
context of mixed categorical survey data. Latent variable modeling
ideas are used, as the observed response is viewed as a categorical
manifestation of a latent continuous variable(s). Several models for
clustering mixed data have been detailed in the literature. Early work
on modeling such data employed the location model
[\citet{lawrence96,hunt99,willse99}], in which the joint distribution of mixed data is
decomposed as the product of the marginal distribution of the
categorical variables and the conditional distribution of the
continuous variables, given the categorical variables. More recently,
\citet{hunt03} reexamined these location models in the presence of
missing data. Latent factor models in particular have generated
interest for modeling mixed data; \citet{quinn04} uses such models in a
political science context. \citet{gruhl13} and \citet{murray13} use factor
analytic models based on a Gaussian copula as a model for mixed data,
but not in a clustering context. \citet{everitt88,everitt90} and \citet
{muthen99} provide an early view of clustering mixed data, including
the use of latent variable models. \citet{cai11}, \citet{browne12} and
\citet{gollini2013} propose clustering models for categorical data based
on a latent variable. However, none of the existing suite of clustering
methods for mixed categorical data has the capability of modeling the
exact nature of the binary, ordinal and nominal variables in the
Agincourt survey data, or the desirable feature of modeling all the
survey items in a unified framework. The clustering model proposed here
presents a unifying latent variable framework by elegantly combining
ideas from item response theory (IRT) and from factor analysis models
for nominal data.

Item response modeling is an established method for analyzing binary or
ordinal response data. First introduced by \citet{thurstone25}, IRT has
its roots in educational testing. Many authors have contributed to the
expansion of this theory since then, including \citet{lord52,rasch60},
\citet{lord68} and \citet{vermunt01}. Extensions include the graded
response model [\citet{samejima69}] and the partial credit model [\citet
{masters82}]. Bayesian approaches to fitting such models are detailed in
\citet{johnson99} and \citet{fox10}. IRT models assume that each observed
ordinal response is a manifestation of a latent continuous variable.
The observed response will
be level $k$, say, if the latent continuous variable lies within a
specific interval. Further, IRT models assume that the latent
continuous variable is a function of both a respondent specific latent
trait variable and item specific parameters.

Modeling nominal response data is typically more complex than modeling
binary or ordinal data, as the set of possible responses is unordered.
A popular model for nominal choice data is the multinomial probit (MNP)
model [\citet{geweke94}]. Bayesian approaches to fitting the MNP model
have been
proposed by \citet{albert93,McCulloch94,nobile98} and \citet{chib98}.
The model has also been extended to include multivariate nominal
responses by \citet{zhang08}. The MNP model treats nominal response data
as a manifestation of an underlying multidimensional continuous latent
variable, which depends on a respondent's covariate information and
some item specific parameters. Here a factor analysis model for nominal
data, similar in nature to the MNP model, is proposed where the
observed nominal response is a manifestation of the multidimensional
latent variable which is itself modeled as a function of both a
respondent's latent trait variable and some item specific parameters.

The structural similarities between IRT models and the MNP model
suggest a hybrid model would be advantageous. Both models have a latent
variable structure underlying the observed data which exhibits
dependence on item specific parameters. Further, the latent variable in
both models has an underlying factor analytic structure through the
dependency on the latent trait. This similarity is exploited and the
models are combined to produce a hybrid model capable of modeling mixed
categorical data types. This hybrid model can be thought of as a factor
analysis model for mixed data.

As stated, the motivation here is the need to substantively explore
clusters of Agincourt households based on mixed categorical survey
data. A~model-based approach to clustering is proposed, in that a
mixture modeling framework provides the clustering machinery.
Specifically, a mixture of the factor analytic models for mixed data is
considered to provide clustering capabilities within the context of
mixed binary, ordinal and nominal response data. The resulting model is
termed the mixture of factor analyzers for mixed data (MFA-MD).

The paper proceeds as follows. Background information about the
Agincourt region of South Africa as well as the socio-economic status
(SES) survey and resulting data set are introduced in Section~\ref{sec:data}. IRT models, a model for nominal response data and the
amalgamation and extension of these models to a MFA-MD model are
considered in Section~\ref{sec:model}. Section~\ref{sec:estimate} is
concerned with Bayesian
model
estimation and inference. The results from fitting the model to the
Agincourt data are presented in Section~\ref{sec:SESresults}. Finally,
discussion of the results and future research areas takes place in
Section~\ref{sec:discuss}.

\section{The Agincourt HDSS data set}
\label{sec:data}

The Health and Demographic Survey System (HDSS) covers an area of
$420$~km$^2$ consisting of 21 villages with a total population of
approximately $82\mbox{,}000$ people. The
infrastructure in the area is mixed. The roads in and surrounding the
study area are in the process of rapidly being upgraded from dirt to
tar. The cost of electricity
is prohibitively high for many households, though it is available in
all villages. A dam has been constructed nearby, but to date there is
no piped
water to dwellings and sanitation is rudimentary. The soil in the area
is generally suited to game farming and there is virtually no commercial
farming activity. Most households contain wage earners who purchase
maize and other foods which they then supplement with home-grown crops
and collected wild foodstuffs.

To explore the SES landscape in Agincourt, data describing assets of
households in the Agincourt study area are analyzed. The data consist
of the responses of $N = 17\mbox{,}617$ households
to each of $J = 28$ categorical survey items. There are $22$ binary
items, $3$ ordinal items and $3$ nominal items. The binary items are
asset ownership indicators for the most part. These items record
whether or not a household owns a particular asset (e.g., whether or
not they own a working car). An example of an ordinal item is the type
of toilet the household uses. This follows an ordinal scale from no
toilet at all to a modern flush toilet. Finally, the power used for
cooking is an example of a nominal item. The household may use
electricity, bottled gas or wood, among others. This is an unordered
set. A full list of survey items is given in Appendix~\ref{app:items}.
For more information on the Agincourt HDSS and on data collection see
\surl{www.agincourt.co.za}.

Previous analyses of similar mixed categorical asset survey data derive
SES strata using principal components analysis. Typically households
are grouped into predetermined categories based on the first principal
scores, reflecting different SES levels
[\citet{vyas06,filmer01,mckenzie05,gwatkin07}]. \citet{filmer01}, for example, examine the
relationship between educational enrollment and wealth in India by
constructing an SES asset index based on principal component scores.
Percentiles are then used to partition the observations into groups
rather than the model-based approach suggested here. In a previous
analysis of the Agincourt HDSS survey data, \citet{collinson09}
construct an asset index for each household. How migration impacts upon
this index is then analyzed, rather than the exploration of SES
considered here. The routine approach of principal components analysis
does not explicitly recognize the data as categorical and, further, the
use of such a one-dimensional index will often miss the natural
groups that exist with respect to the whole collection of assets and
other possible SES variables. The model proposed here aims to alleviate
such issues.

\section{A mixture of factor analyzers model for mixed data}
\label{sec:model}
A mixture of factor analyzers model for mixed data (MFA-MD) is proposed
to explore SES clusters of Agincourt households. Each component of the
MFA-MD model is a hybrid of an IRT model and a factor analytic model
for nominal data. In this section IRT models for ordinal data and a
latent variable model for nominal data are introduced, before they are
combined and extended to the MFA-MD model.

\subsection{Item response theory models for ordinal data}
\label{subsec:IRTord}

Suppose item $j$\break  (for~$j = 1, \ldots, J$) is ordinal and the set of
possible responses is denoted $\{1, 2, \ldots, K_j\}$, where $K_j$
denotes the number of response levels to item $j$. IRT models assume
that, for respondent $i$, a latent Gaussian variable $z_{ij}$
corresponds\vadjust{\goodbreak} to each categorical response $y_{ij}$. A Gaussian link
function is assumed, though other link functions, such as the logit,
are detailed in the IRT literature [\citet{fox10,lord68}].

For each ordinal item $j$ there exists a vector of threshold parameters
$\underline{\gamma}_j =  ( \gamma_{j, 0}, \gamma_{j, 1}, \ldots,
\gamma_{j, K_j}  )$, the elements of which are constrained such that
\[
- \infty= \gamma_{j,0} \leq\gamma_{j,1} \leq\cdots\leq\gamma
_{j,K_j} = \infty.
\]
For identifiability reasons [\citet{albert93,quinn04}] $\gamma_{j,1} =
0$. The observed ordinal response, $y_{ij}$, for respondent $i$ is a
manifestation of the latent variable $z_{ij}$, that is,
%
\begin{equation}
\label{eqn:ordinalcutoffs} \mbox{ if } \gamma_{j,k-1} \leq z_{ij} \leq
\gamma_{j,k} \qquad\mbox{then }  y_{ij} = k.
\end{equation}
That is, if the underlying latent continuous variable lies within an
interval bounded by
the threshold parameters $\gamma_{j,k-1}$ and $\gamma_{j,k}$, then the
observed ordinal response is level $k$.

In a standard IRT model, a factor analytic model is then used to model
the underlying latent variable $z_{ij}$. It is assumed that the mean of
the conditional distribution of $z_{ij}$ depends on a $q$-dimensional,
respondent specific, latent variable $\underline{\theta}_i$ and on some
item specific parameters. The latent variable $\underline{\theta}_{i}$
is sometimes referred to as the latent trait or a respondent's ability
parameter in IRT. Specifically, the underlying latent variable $z_{ij}$
for respondent $i$ and item $j$ is assumed to be distributed as
\[
z_{ij}| \underline{\theta}_i \sim N\bigl(
\mu_j + \underline{\lambda }_j^T\underline{
\theta}_i, 1\bigr).
\]
The parameters $\underline{\lambda}_j$ and $\mu_j$ are usually termed
the item discrimination parameters and the negative item difficulty
parameter, respectively. As in \citet{albert93}, a probit link function
is used so the variance of $z_{ij}$ is $1$.

Under this model, the conditional probability that a response takes a
certain ordinal value can be expressed as the difference between two
standard Gaussian cumulative distribution functions, that is,
$P(y_{ij}=k |\underline{\lambda}_j, \mu_j, \underline{\gamma}_{j},
\underline{\theta}_i) $ is
%
\begin{equation}
\label{eqn:intract}  \Phi\bigl[\gamma_{j,k} - \bigl( \mu_j +
\underline{\lambda}_j^T\underline{\theta}_i
\bigr)\bigr] - \Phi\bigl[\gamma _{j,k-1} - \bigl(\mu_j +
\underline{\lambda}_j^T\underline{\theta}_i
\bigr)\bigr].
\end{equation}

Since a binary item can be viewed as an ordinal item with two levels (0
and 1, say), the IRT model can also be used to model binary response
data. The threshold parameter for a binary item $j$ is $\underline
{\gamma}_{j} = (-\infty, 0, \infty)$ and, hence,
\[
P(y_{ij}=1 |\underline{\lambda}_j, \mu_j,
\underline{\gamma}_{j}, \underline{\theta}_i) = \Phi
\bigl(\mu_j + \underline{\lambda }_j^T
\underline{\theta}_i \bigr).
\]

\subsection{A factor analytic model for nominal data}
\label{subsec:IRTnom}

Modeling nominal response data is typically more complicated than
modeling ordinal data since the set of possible responses is no longer
ordered. The set of nominal responses for item $j$ is
denoted $\{1, 2, \ldots, K_j\}$ such that 1 corresponds to the first
response choice while $K_j$ corresponds to the last response choice,
but where no inherent ordering among the choices is assumed.

As detailed in Section~\ref{subsec:IRTord}, the IRT model for ordinal
data posits a one-dimensional latent variable for each observed ordinal
response. In the factor analytic model for nominal data proposed here,
a $K_j - 1$-dimensional latent variable is required for each observed
nominal response. That is, the latent variable for observation $i$
corresponding to nominal item $j$ is denoted
\[
\underline{z}_{ij} = \bigl(z_{ij}^1, \ldots,
z_{ij}^{K_j - 1}\bigr).
\]
The observed nominal response is then assumed to be a manifestation of
the values of the elements of $\underline{z}_{ij}$ relative to each
other and to a cutoff point, assumed to be $0$. That is,
\[
y_{ij} = \cases{ 1, &\quad $\mbox{if $ \max
_k\bigl\{z_{ij}^k\bigr\} < 0$};$
\vspace*{2pt}\cr
k, & \quad $\mbox{if $z_{ij}^{k-1} = \max_k\bigl
\{z_{ij}^k\bigr\}$ and $z_{ij}^{k-1} > 0$
for $k = 2, \ldots, K_{j}$}.$}
\]

Similar to the IRT model, the latent vector $\underline{z}_{ij}$ is
modeled via a factor analytic model. The mean of the conditional
distribution of $\underline{z}_{ij}$ depends on a respondent specific,
$q$-dimensional, latent trait, $\underline{\theta}_i$, and item
specific parameters, that is,
$\underline{z}_{ij}| \underline{\theta}_i \sim\operatorname
{MVN}_{K_{j}-1}(\underline{\mu}{}_j + \Lambda_j \underline{\theta}_i,
\mathbf{I})$,
where $\mathbf{I}$ denotes the identity matrix. The loadings matrix
$\Lambda_{j}$ is a $(K_{j} - 1) \times q$ matrix, analogous to the item
discrimination parameter in the IRT model of Section~\ref{subsec:IRTord}; likewise, the mean
$\underline{\mu}{}_{j}$ is analogous
to the item difficulty parameter in the IRT model.

It should be noted that binary data could be regarded as either ordinal
or nominal. The model proposed here is equivalent to the model proposed
in Section~\ref{subsec:IRTord} when $K_j = 2$.

\subsection{A factor analysis model for mixed data}
\label{subsec:Hyb}

It is clear that the IRT model for ordinal data (Section~\ref{subsec:IRTord}) and the factor analytic model for nominal data
(Section~\ref{subsec:IRTnom}) are similar in structure. Both model the
observed data as a manifestation of an underlying latent variable,
which is itself modeled using a factor analytic structure. This
similarity is exploited to obtain a hybrid factor analysis model for
mixed binary, ordinal and nominal data.

Suppose $Y$, an $N \times J$ matrix of mixed data, denotes the data
from $N$ respondents to $J$ survey items. Let $O$ denote the number of
binary items plus the number of ordinal items, leaving $J - O$ nominal
items. Without loss of generality, suppose that the binary and ordinal
items are in
the first $O$ columns of $Y$ while the nominal items are in the
remaining columns.

The binary and ordinal items are modeled using an IRT model and the
nominal items using the factor analytic model for nominal data.
Therefore, for each respondent $i$ there are $O$ latent continuous
variables corresponding to the ordinal items and $J-O$ latent
continuous vectors corresponding to the nominal items. The latent
variables and latent vectors for respondent $i$ are collected together
in a single $D$-dimensional vector $\underline{z}_{i}$, where $D = O +
\sum_{j=O+1}^{J}(K_j - 1)$. That is, underlying respondent $i$'s set of
$J$ binary, ordinal and nominal responses lies the latent vector
\[
\underline{z}_i = \bigl(z_{i1}, \ldots, z_{iO},
\ldots, z_{iJ}^1, \ldots, z_{iJ}^{K_J-1}
\bigr).
\]
This latent vector is then modeled using a factor analytic structure:
%
\begin{equation}
\label{eqn:FA-MD} \underline{z}_{i}| \underline{\theta}_i
\sim\operatorname {MVN}_{D}(\underline{\mu} + \Lambda\underline{
\theta}_i, \mathbf{I}).
\end{equation}
The $D \times q$-dimensional matrix $\Lambda$ is termed the loadings
matrix and $\underline{\mu}$ is the $D$-dimensional mean vector.
Combining the IRT and factor analytic models in this way facilitates
the modeling of binary, ordinal and nominal response data in an elegant
and unifying latent variable framework.

The model in (\ref{eqn:FA-MD}) provides a parsimonious factor analysis
model for the high-dimensional latent vector $\underline{z}_{i}$ which
underlies the observed mixed data. As in any model which relies on a
factor analytic structure, the loadings matrix details the relationship
between the low-dimensional latent trait $\underline{\theta}_{i}$ and
the high-dimensional latent vector $\underline{z}_{i}$. Marginally, the
latent vector is distributed as
\[
\underline{z}_{i} \sim\operatorname{MVN}_{D}\bigl(\underline{\mu},
\Lambda\Lambda ^{T} + \mathbf{I}\bigr),
\]
resulting in a parsimonious covariance structure for $\underline{z}_{i}$.

\subsection{A mixture of factor analyzers model for mixed data}
\label{subsec:HybMix}

To facilitate clustering when the observed data are mixed categorical
variables, a mixture modeling framework can be imposed on the hybrid
model defined in Section~\ref{subsec:Hyb}. The resulting model is
termed the mixture of factor analyzers model for mixed data. In the
MFA-MD model, the clustering is deemed to occur at the latent variable
level, that is, under the MFA-MD model the distribution of the latent
data $\underline{z}_i$ is modeled as a mixture of $G$ Gaussian densities
%
\begin{equation}
\label{eqn:MFA-MD} f(\underline{z}_i)  =  \sum
_{g=1}^{G}\pi_g \operatorname{MVN}_{D}
\bigl( \underline{\mu}{}_g, \Lambda_g
\Lambda_g^T +\mathbf{I}_D \bigr).
\end{equation}
The probability of belonging to cluster $g$ is denoted by $\pi_g$,
where $\sum_{g=1}^{G}\pi_g = 1$ and $\pi_g > 0\ \forall g$. The mean
and loading parameters are cluster specific.

As is standard in a model-based approach to clustering
[\citet{fraley98,celeux00}], a latent indicator variable, $\underline{\ell}_i = (\ell
_{i1}, \ldots, \ell_{iG})$, is introduced for each respondent $i$. This
binary vector indicates the cluster to which individual $i$ belongs,
that is, $l_{ig} = 1$ if $i$ belongs to cluster~$g$; all other entries
in the vector are $0$.\vadjust{\goodbreak} Under the model in (\ref{eqn:MFA-MD}), the
augmented likelihood function for the $N$ respondents is then given by
%
\begin{eqnarray}\label{eqn:like}
&&
\mathcal{L}(\underline{\pi}, \tilde{\Lambda}, \Gamma, Z, \Theta, L | Y)
\nonumber\\
&&\qquad = \prod_{i=1}^{N} \prod
_{g=1}^{G} \Biggl\{ \pi_g \Biggl[ \prod
_{j=1}^{O} \prod
_{k=1}^{K_j} N\bigl(z_{ij} |\tilde{
\underline{\lambda}}_{gj}^{T} \tilde{\underline{
\theta}}_{i}, 1\bigr)^{\mathbb{I}
\{\gamma_{j,k-1} < z_{ij} < \gamma_{j,k} | y_{ij}\}} \Biggr]
\\
 &&\hspace*{36pt}\qquad\quad{}  \times \Biggl[\prod_{j=O+1}^{J}
\prod_{k=2}^{K_j} \prod
_{s=1}^{3} N\bigl(z_{ij}^{k-1} |
\tilde{\underline{\lambda}}{}_{gj}^{k-1^{T}} \tilde{\underline{\theta
}}_{i}, 1\bigr)^{\mathbb{I}(\mathrm{case}\  s|y_{ij})}
 \Biggr] \Biggr\}^{\ell_{ig}},\nonumber
\end{eqnarray}
where $\tilde{\underline{\theta}}_i = (1, \theta_{i1}, \ldots, \theta
_{iq})^T$ and $\tilde{\Lambda}_g$ is the matrix resulting from the
combination of
$\underline{\mu}{}_g$ and $\Lambda_g$ so that the first column of $\tilde
{\Lambda}_g$ is $\underline{\mu}{}_g$. Thus, the $d${th} row of
$\tilde{\Lambda}_g$ is $\tilde{\underline{\lambda}}_{gd} = (\mu_{gd},
\lambda_{gd1}, \ldots, \lambda_{gdq})$.

The likelihood function in (\ref{eqn:like}) depends on the observed
responses $Y$ through the indicator functions. In the ordinal part of
the model, the observed $y_{ij}$ restricts the interval in which
$z_{ij}$ lies, as detailed in (\ref{eqn:ordinalcutoffs}). In the
nominal part of the model, $z_{ij}^{k-1}$ is restricted in one of three
ways, depending on the observed $y_{ij}$. The three cases $\mathbb
{I}(\mathrm{case}\  s|y_{ij})$ for $s = 1, 2, 3$ are defined as follows:
\begin{itemize}
\item$\mathbb{I}(\mathrm{case}\  1|y_{ij}) = 1$ if $y_{ij} = 1$, that is,
$ \max_k\{z_{ij}^k\} < 0$.
\item$\mathbb{I}(\mathrm{case}\  2|y_{ij}) = 1$ if $ y_{ij} = k$, that is,
$z_{ij}^{k-1} =  \max_k\{z_{ij}^k\}$ and $z_{ij}^{k-1} > 0$.
\item$\mathbb{I}(\mathrm{case}\  3|y_{ij}) = 1$ if $ y_{ij} \neq 1 \land
y_{ij} \neq k$, that is, $z_{ij}^{k-1} <  \max_k\{z_{ij}^k\}$.
\end{itemize}
An example of how this latent variable formulation gives rise to
particular nominal responses is given in Appendix \ref{app:toy}.

The MFA-MD model proposed here is related to the mixture of factor
analyzers model [\citet{ghahramani97}] which is appropriate when the
observed data are continuous in nature. \citet{fokoue03} detail a
Bayesian treatment of such a model; \citet{mcnicholas08} detail a suite
of parsimonious mixture of factor analyzer models.

The MFA-MD model developed here provides a novel approach to clustering
the mixed data in the Agincourt survey in a unified framework. In
particular, the MFA-MD model has two novel features: (i) it has the
capability to appropriately model the exact nature of the data in the
Agincourt survey, in particular, the nominal data, and (ii) it has the
capability of modeling all the survey items in a unified manner.

\section{Bayesian model estimation}
\label{sec:estimate}
A Bayesian approach using Markov chain Monte Carlo (MCMC) is utilized
for fitting the MFA-MD model to the Agincourt survey data. Interest
lies in the cluster membership vectors $L$ and the mixing proportions
$\underline{\pi}$, and in the underlying latent variables $Z$, the
latent traits $\Theta$, the item parameters $\tilde{\Lambda}_{g}
(\forall g = 1, \ldots, G$) and the threshold parameters $\Gamma$.

\subsection{Prior and posterior distributions}
\label{subsec:priors}

To fit the MFA-MD model in a Bayesian framework, prior distributions
are required for all unknown parameters. As in \citet{albert93}, a
uniform prior is specified for the threshold parameters. Conjugate
prior distributions are specified for the other model parameters:
\[
p(\tilde{\underline{\lambda}}_{gd}) = \operatorname{MVN}_{(q+1)}(
\underline{\mu }_{\lambda},  \Sigma_{\lambda}), \qquad p(\underline{\pi}) =
\operatorname{Dirichlet}(\underline{\alpha}).
\]
In terms of latent variables, it is assumed the latent traits
$\underline{\theta}_i$ follow a standard multivariate Gaussian
distribution while the latent indicator variables, $\underline{\ell
}_i$, follow a $\operatorname{Multinomial}(1, \underline{\pi})$ distribution. Further,
conditional on membership of cluster $g$, the latent variable
$\underline{z}_{i}| l_{ig} = 1 \sim\operatorname{MVN}_{D}(\underline{\mu}{}_{g},
\Lambda_{g} \Lambda_{g}^{T} + \mathbf{I})$. Combining these latent
variable distributions and prior distributions with the likelihood
function specified in (\ref{eqn:like}) results in the joint posterior
distribution, from which samples of the model parameters and latent
variables are drawn using a MCMC sampling scheme.

\subsection{Estimation via a Markov chain Monte Carlo sampling scheme}

As the marginal distributions of the model parameters cannot be
obtained analytically, a MCMC sampling scheme is employed. All
parameters and latent variables are sampled using Gibbs sampling, with
the exception of the threshold parameters $\Gamma$, which are sampled
using a Metropolis--Hastings step.

The full conditional distributions for the latent variables and model
parameters are detailed below; full derivations are given in \citet
{mcparland2014}:

\begin{itemize}
\item Allocation vectors. For $i=1, \ldots, N$: $\underline{\ell
_i}|\cdots\sim\operatorname{Multinomial}(\underline{p})$, where $\underline
{p}$ is defined in \citet{mcparland2014}.

\item Latent traits. For $i=1, \ldots, N$:
$\underline{\theta}_i| \cdots\sim\operatorname{MVN}_q \{  [ \Lambda
_g^T \Lambda_g + \mathbf{I}  ]^{-1} [
\Lambda_g^T (\underline{z}_i - \underline{\mu}{}_g )  ],\break  [ \Lambda_g^T \Lambda_g + \mathbf{I}
]^{-1} \}$.\vspace*{1.5pt}

\item Mixing proportions:
$\underline{\pi} | \cdots\sim\operatorname{Dirichlet}(n_1 + \alpha_1, \ldots,
n_g + \alpha_G)$ where $n_g = \sum_{i=1}^{N}\ell_{ig}$.

\item Item parameters. For $g=1, \ldots, G$ and $d=1, \ldots, D$:
$\tilde{\underline{\lambda}}_{gd} | \cdots\sim\break \operatorname{MVN}_{(q+1)}
\{  [ \Sigma_{\lambda}^{-1} +
\tilde{\Theta}_g^{T}\tilde{\Theta}_g  ]^{-1} [\tilde{\Theta
}_g^{T}\underline{z}_{gd} + \Sigma_{\lambda}^{-1}\underline{\mu
}_{\lambda}
],  [ \Sigma_{\lambda}^{-1} + \tilde{\Theta}_g^{T}\tilde
{\Theta}_g  ]^{-1}  \}$, where\break  $\underline{z}_{gd} = \{
z_{id}\}$ for
all respondents $i$ in cluster $g$ and $\tilde{\Theta}_g$ is a matrix,
the rows of which are $\tilde{\underline{\theta}}_i$ for members of
cluster $g$.
\end{itemize}
The full conditional distribution for the underlying latent data $Z$
follows a truncated Gaussian distribution. The point(s) of truncation
depends on the nature of the corresponding item, the observed response
and the values of $Z$ from the previous iteration of the MCMC chain.
The distributions are truncated to satisfy the conditions detailed in
Section~\ref{sec:model}. Thus, the latent data $Z$ are updated as follows:
\begin{itemize}
\item If item $j$ is ordinal and $y_{ij}=k$, then
\[
z_{ij}| \cdots\sim N^T \bigl(\tilde{\underline{
\lambda}}{}_{gj}^T\tilde {\underline{\theta}}_i,
1 \bigr),
\]
where the distribution is truncated on the interval $(\gamma_{j,k-1},
\gamma_{j,k})$.

\item If item $j$ is nominal, then
\[
z_{ij}^k | \cdots\sim N^T \bigl( \tilde{
\underline{\lambda }}_{gj}^{k^T}\tilde{\underline{
\theta}}_i, 1 \bigr),
\]
where $\tilde{\underline{\lambda}}{}_{gj}^k$ is the row of $\tilde{\Lambda
}_g$ corresponding to $z_{ij}^k$ and the truncation intervals are
defined as follows:
\begin{itemize}[$-$]
\item[$-$] if $y_{ij}=1$, then $z_{ij}^k \in(-\infty,0)$ for $k = 1, \ldots, K_{j}-1$.
\item[$-$] if $y_{ij}=k > 1$, then:
\begin{enumerate}
\item$z_{ij}^{k-1} \in(\tau, \infty)$ where $\tau= \max (0,
\max_{l \neq k-1}\{z_{ij}^l\} )$.
\item for $l = 1, \ldots, k-2, k, \ldots, K_{j}-1$ then $z_{ij}^{l} \in
( - \infty, z_{ij}^{k-1} )$.
\end{enumerate}
\end{itemize}
\end{itemize}
Note that, in the case of $y_{ij} = k >1$ above, the values $
z_{ij}^{l} $ considered in the evaluation of $\tau$ in step 1 are those
from the previous point in the MCMC chain. The value of $z_{ij}^{k-1}$
in step 2 is that sampled in step 1.

As a uniform prior is specified for the threshold parameters, the
posterior full conditional distribution of $\underline{\gamma}_j$ is
also uniform, facilitating the use of a Gibbs sampler. However, if
there are large numbers of observations in adjacent response
categories, very slow mixing may be observed. Thus, as in \citet
{cowles96,fox10,johnson99}, a Metropolis--Hastings step is used to
sample the threshold parameters; the overall sampling scheme employed
is therefore a Metropolis-within-Gibbs sampler.

Briefly, the Metropolis--Hastings step involves proposing candidate
values $v_{j,k}$ (for $k = 2, \ldots, K_{j}-1$) for $\gamma_{j,k}$ from
the Gaussian distribution $\mathrm{N}^T(\gamma_{j,k}^{(t-1)},\break \sigma
_{\mathrm{MH}}^2)$ truncated to the interval $(v_{j,k-1}, \gamma
_{j,k+1}^{(t-1)})$,\vspace*{1pt} where $\gamma_{j,k+1}^{(t-1)}$ is the value of
$\gamma_{j,k+1}$ sampled at iteration $(t-1)$. The threshold vector
$\underline{\gamma}_j$ is set equal to the proposed vector, $\underline
{v}_j$, with probability $\beta= \min(1, R)$, where $R$ is defined in
\citet{mcparland2014}.
The tuning parameter $\sigma_{\mathrm{MH}}^2$ is selected to achieve appropriate
acceptance rates.

This Metropolis-within-Gibbs sampling scheme is iterated until
convergence, after which the samples drawn are from the joint posterior
distribution of all the model parameters and latent variables of the
MFA-MD model.

\subsection{Model identifiability}
\label{sec:identify}

The MFA-MD model as described is not identifiable. One identifiability
aspect of the model concerns the threshold parameters. If a constant is
added to the threshold parameters for an ordinal item $j$ and the same
constant is added\vadjust{\goodbreak} to the corresponding mean parameter(s), the
likelihood remains unchanged. Therefore, as outlined in Section~\ref{subsec:IRTord}, the second element $\gamma_{j1}$ of the vector of
threshold parameters, $\underline{\gamma}_j$, is fixed at $0$ for all
ordinal items $j$.

The model is also rotationally invariant due to its factor analytic
structure. Many approaches to this identifiability issue have been
proposed in the literature. A popular solution is that proposed by \citet
{geweke96} where the loadings matrix is constrained such that the first
$q$ rows have a lower triangular form and the diagonal elements are
positive. This approach is adopted by \citet{quinn04} and \citet
{fokoue03}, among others. However, this approach enforces an ordering
on the variables [\citet{aguilar00}] which is not appropriate under the
MFA-MD model.

Here, the approach to identifying the MFA-MD model is based on that
suggested by \citet{hoff02} and \citet{handcock07} in relation to latent
space models for network data. Instead of imposing a particular form on
the loadings matrices, the MCMC samples are post-processed using
Procrustean methods. Each sampled $\Lambda_g$ is rotated and/or
reflected to match as closely as possible to a reference loadings
matrix. The latent traits, $\underline{\theta}_i$, are then subjected
to the same transformation. The sample mean of these transformed values
is then used to estimate the mean of the posterior distribution.

Conditional on the cluster memberships on convergence of the MCMC
chain, a~factor analysis model is fitted to the underlying latent data
within each cluster. The estimated loadings matrix obtained is used as
the reference matrix for each cluster. Only the saved MCMC samples need
to be subjected to this transformation, which is done post hoc and is
computationally cheap.

\section{Results: Fitting the MFA-MD model to the Agincourt data}
\label{sec:SESresults}

In order to describe and understand the SES landscape in the Agincourt
region, the MFA-MD model is fitted to the asset survey data. Varying
the number of clusters $G$ and the dimension of the latent trait $q$
allows consideration of a wide range of MFA-MD models. Choosing the
optimal MFA-MD model is difficult, as likelihood based criteria, such
as the Bayesian Information Criterion or marginal likelihood
approaches, are not available since the likelihood cannot be evaluated.
However, within the sociological setting in which the MFA-MD is applied
here, the existence of SES clusters is well motivated
[\citet{weeden12,erikson92,svalfors06,alkema08}]. Further, the literature suggests
small numbers ($\approx$ 3) of such SES clusters typically exist.
Hence, to examine the (dis)similar features of the SES clusters in the
Agincourt region, MFA-MD models with $G = 2, \ldots, 6$ and $q = 1, 2$
are fitted to the data. Models in which $q > 2$ were not considered for
reasons of parsimony.

Trace plots of the Markov chains were used to judge convergence and
examples are presented in Appendix \ref{app:cvg}. To achieve
satisfactory mixing in the Metropolis--Hastings\vadjust{\goodbreak} sampling of the
threshold parameters, $\underline{\gamma}_j$, a small proposal variance
was required. Acceptance rates of 20--30\% were observed. The Jeffreys
prior, $\operatorname{Dirichlet}(\underline{\alpha} = \frac{1}{2} \mathbf{1})$, was
specified for the mixing weights $\underline{\pi}$. A~multivariate
normal prior with mean $\underline{\mu}{}_{\lambda} = \mathbf{0}$ and
covariance matrix $\Sigma_{\lambda} = 5\mathbf{I}$ was specified for
$\underline{\tilde{\lambda}}_{gd}$. In the absence of strong prior
information, this relatively uninformative prior was chosen. It should
be noted, however, that flat priors can lead to improper posterior
distributions in the context of mixture models [\citet{fruhwirth2006}].
To assess prior sensitivity, different values for the hyperparameters
were trialled, namely, $\mu_\lambda\in\{0, 0.5\}$ and $\Sigma_\lambda
\in\{\mathbf{I}, 1.25\mathbf{I}, 2.5\mathbf{I}, 5\mathbf{I}\}$. All
hyperparameter values produced similar substantive clustering results,
indicating that prior sensitivity does not appear to be an issue,
however, a more thorough exploration may prove informative. The label
switching problem was addressed using methods detailed in \citet{stephens00}.

\subsection{Model assessment}
\label{subsec:ModFit}

Given the question of interest [i.e., what are the (dis)similar
features of the SES clusters in the set of Agincourt households?], and
due to the unavailability of a formal model selection criterion for the
MFA-MD model, focus is placed on models which are substantively
interesting and fit well. Model fit is assessed in an exploratory
manner using three established statistical tools: posterior predictive
checks, clustering uncertainty and residual analysis.

\subsubsection{Posterior predictive checks}
\label{subsec:postpred}

A natural approach to assessing model fit within the Bayesian paradigm
is via posterior predictive model checking [\citet{gelman2003}].
Replicated data are simulated from the posterior predictive
distribution and compared to the observed data. Given the multivariate
and discrete nature of the observed survey data, a discrepancy measure
which focuses on response patterns across the set of assets is employed
to compare observed and replicated data. \citet{erosheva2007} and \citet
{gollini2013} employ truncated sum of squared Pearson residuals (tSSPR)
to assess model fit in the context of clustering categorical data. The
standard SSPR examines deviations between observed and expected counts
of response patterns; the truncated SSPR evaluates the SSPR only for
the $T$ most frequently observed response patterns.

\begin{figure}

\includegraphics{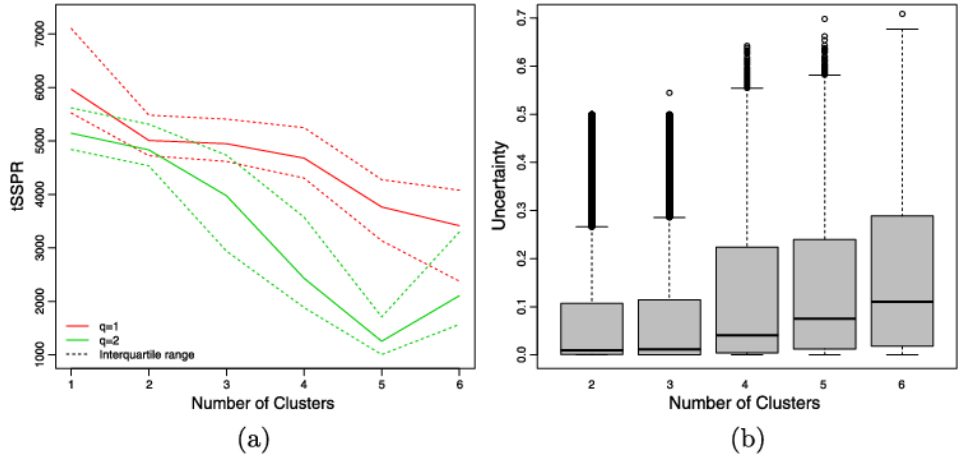}

\caption{Assessing model fit.
\textup{(a)} The median tSSPR, and its associated uncertainty, across a
range of MFA-MD models. \textup{(b)}
Box plots of clustering uncertainty across models with
between 2 and 6 clusters, and a 1-dimensional
latent trait.}
\label{fig:ModelFit}
\end{figure}

In the MFA-MD setting, however, computing expected counts is
intractable since this involves evaluating response pattern
probabilities, which requires integrating a multidimensional truncated
Gaussian distribution, where truncation limits differ and are dependent
across the dimensions. Hence, here posterior predictive data are used
to obtain a pseudo tSSPR. Replicated data sets $\mathbf{Y}_r$ for $r =
1, \ldots, R$ are simulated from the posterior predictive distribution
and for each the tSSPR is computed where
\[
\mathit{tSSPR}_{r} = \sum_{t=1}^{T}
\frac{(o_{t} - p_{t})^2}{p_{t}}.\vadjust{\goodbreak}
\]
Here $o_{t} = $ observed count of response pattern $t$ and $ p_{t} = $
predicted count of response pattern $t$ in replicated data set $\mathbf
{Y}_{r}$. Response patterns observed 30 times or more are considered
here, which is equivalent to a truncation level of $T = 20$. This
measure is computed for $R = 1500$ replicated data sets across MFA-MD
models with $G = 1, \dots, 6$ and $q = 1, 2$. The $G = 1$ case is
included for completion. The median of the $R$ tSSPR values for each
model considered is illustrated in Figure~\ref{fig:ModelFit}(a), along
with the quantile based interquartile range.

Based on the median tSSPR values, the improvement in fit from $q=1$ to
$q=2$ across $G$ was felt to be insufficient to substantiate focusing
on the $q = 2$ models, given the reduction in parsimony. Examination of
the parameters of the $q=2$ model for a fixed $G$ also provided little
substantive insight over the $q=1$ model. Models with $G = 2$, $G = 3$
and $G = 4$ (with $q=1$) all seem to fit equivalently well; this
observation is also apparent under other truncation levels $T$, as
illustrated in \citet{mcparland2014}. Further, the median tSSPR values
support the literature's assertion that SES clusters exist, that is,
that $G > 1$.

\subsubsection{Clustering uncertainty}
\label{subsec:uncertainty}

Clustering uncertainty [\citet{bensmail97,gormley06}] is an exploratory
tool which helps assess models in the context of clustering. The
uncertainty with which household $i$ is assigned to its cluster may be
estimated by
\[
U_i = \min_{g=1,\ldots, G} \bigl\{1 - \PP(\mbox{cluster } g  |  \mbox{household }i)\bigr\}.
\]
If household $i$ is strongly associated with cluster $g$, then $U_i$
will be small.

Box plots of the clustering uncertainty of each household under models
with $G = 2, \ldots, 6$ (and $q = 1$) are shown in Figure~\ref{fig:ModelFit}(b). The uncertainty values are low in general,
indicating that households are assigned to clusters with a high degree
of confidence. Low values are observed for the $G = 2$ and $G = 3$
models, with a notable increase for higher numbers of clusters.

\begin{figure}

\includegraphics{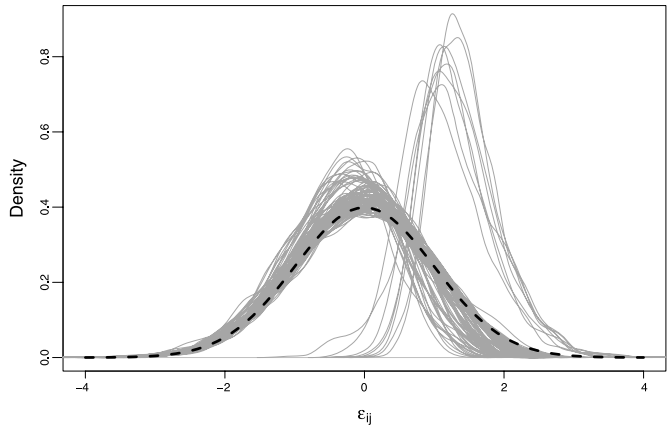}

\caption{Bayesian latent residuals, corresponding to the \emph{cattle}
survey item, for 100 randomly sampled households under the $G=3$ model
with a 1-dimensional latent trait. The dashed black line is the
standard normal curve.}\label{fig:CattleRes}  
\end{figure}

\subsubsection{Bayesian latent residuals analysis}
\label{subsec:residuals}

The posterior predictive checks and the clustering uncertainties
suggest that models with $G=2$, $G=3$ and $G=4$ (and $q=1)$ appear to
fit well and are relatively parsimonious. Focus is given to these
models, and Bayesian latent residuals [\citet{johnson99,fox10}] are
employed to investigate their model fit. Bayesian latent residuals,
defined by $\varepsilon_{ij} = z_{ij} - \tilde{\underline{\lambda}}{}_{gj}^T
\tilde{\underline{\theta}}_{i}$, should follow a standard normal
distribution.\vspace*{1pt} The Bayesian latent residuals follow their theoretical
distribution reasonably well for the three models under focus; Figure~\ref{fig:CattleRes} shows kernel density estimate curves of the
Bayesian latent residuals corresponding to the \emph{cattle} item for a
random sample of 100 households. The curves are estimated based on the
residuals at each MCMC iteration. Residuals which do not appear to
follow a standard normal distribution correspond to responses which
were unusual given the household's cluster membership.
Further examples of such residual plots are given in \citet{mcparland2014}.

The three approaches to assessing model fit suggest that focus should
be given to models with $G=2$, $G=3$ and $G=4$, and $q=1$. As the $G=3$
and $G=4$ models give deeper insight to the SES structure of the
Agincourt households than the $G=2$ model, the $G=3$ model is explored
in detail in Section~\ref{subsec:3CompRes}; a~substantive comparison
with the $G=4$ model is provided in Section~\ref{subsec:4CompRes}, in
which the $G=2$ model is also discussed.

\subsection{Results: Three-component MFA-MD model}
\label{subsec:3CompRes}

The clustering resulting from fitting a $3$-component MFA-MD model,
with a one-dimensional latent trait, divides the Agincourt households
into $3$ distinct homogeneous subpopulations, with intuitive
socio-economic characteristics.

The conditional probability that household $i$ belongs to cluster $g$
can be estimated from the MCMC samples by dividing the number of times
household $i$ was allocated to group $g$ by the number of samples. A
``hard'' clustering is then obtained by considering $\max_g \PP(\mbox
{cluster } g  |  \mbox{household }i)$, $\forall i$, and assigning
households to the cluster for which this maximum is achieved.

\begin{table}
\def\arraystretch{0.95}
\caption{The cardinality of each group and the modal
response to items on which the modal response differs across groups}
\label{modes3}
\begin{tabular*}{\textwidth}{@{\extracolsep{\fill}}lccc@{}}
\hline
\textit{\textbf{G}} & \textbf{1} & \textbf{2} & \multicolumn{1}{c@{}}{\textbf{3}}\\
\hline
\# & 7864 & 6543 & 3210 \\[3pt]
\# Bedrooms & 2 & 2 & $\leq$1 \\
Separate Living Area & Yes & Yes & No \\
Toilet Facilities & Yard & Yard & Bush \\
Toilet Type & Pit & Pit & None \\
Power for Cooking & Electric & Wood & Wood \\
Stove & Yes & No & No \\
Fridge & Yes & Yes & No \\
Television & Yes & Yes & No \\
Video & Yes & No & No \\
Poultry & No & Yes & No\\
\hline
\end{tabular*}
\end{table}

The modal responses to items for which the modal response differed
across groups are presented in Table~\ref{modes3}.
These statistics only tell part of the story, however, and the
distribution of responses will be analyzed later.

It can be seen from Table~\ref{modes3} that cluster 1 is a
modern/wealthy group of households.
The modal responses indicate that households in this cluster are most
likely to possess modern conveniences such as a stove, a fridge and
also some luxury items such as a television.

In contrast, cluster 3 is a less wealthy group. Households in this
group are likely to have poor sanitary facilities---the modal response
to the location of toilet facilities and the type of toilet are
``bush'' and ``none,'' respectively. Households in cluster 3 are also
less likely to own modern conveniences such as a fridge or
television.

The socio-economic status of cluster 2 is somewhere between that of the
other two groups, but closer to cluster 1 than 3. Households in cluster
2 are likely to have better sewage facilities and larger dwellings than
those in cluster 3 but lack some luxury assets such as a video player.
They are
also likely to keep poultry and cook with wood rather than electricity,
which suggests this group may be less modern than cluster 1 to some degree.

It is interesting to note that the largest group is the wealthy/modern
cluster 1, while the smallest group is cluster 3 who have the lowest
living standards.

An almost identical table to Table~\ref{modes3} was produced for a
3-component model with a 2-dimensional latent trait. There were some
further differences in the Power for Lighting and Cell Phone items but
the clusters have the same substantive interpretation.

\begin{figure}

\includegraphics{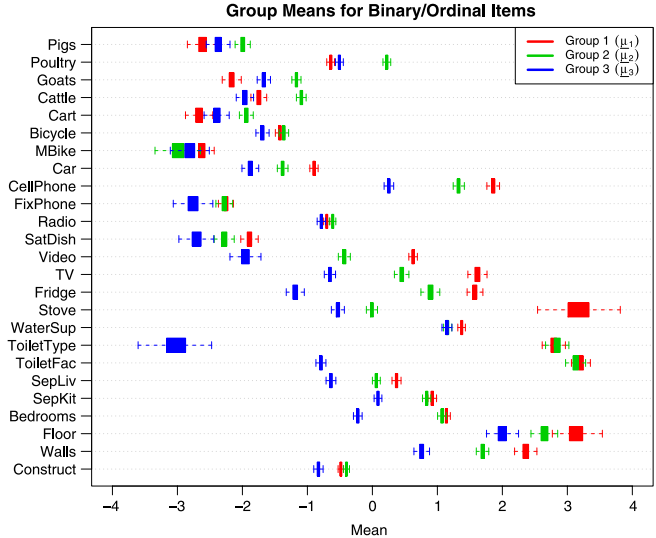}

\caption{Box plots of MCMC samples of the dimensions of the cluster
means, $\underline{\mu}{}_g$, corresponding to binary and ordinal items.}
\label{fig:box_ord}\vspace*{-2pt}
\end{figure}

\begin{figure}

\includegraphics{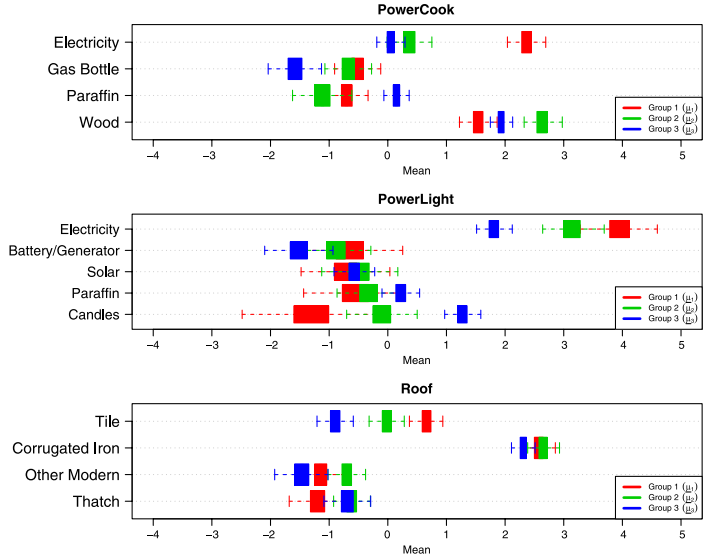}

\caption{Box plots of MCMC samples of the dimensions of the cluster
means, $\underline{\mu}{}_g$, corresponding to nominal items. The first
plot shows box plots of the means of the latent dimensions relating to
the \emph{PowerCook} item, the second shows the means of the dimensions
representing the \emph{PowerLight} item and the third shows the means
of the dimensions corresponding to the \emph{Roof}
item.}\label{fig:box_nom}\vspace*{-2pt}
\end{figure}

A more detailed picture of how the groups differ from each other is
presented in Figures~\ref{fig:box_ord} and~\ref{fig:box_nom}. Box
plots of the MCMC samples of the cluster specific mean parameter
$\underline{\mu}{}_g$ are shown in these figures. The box plots for the
binary/ordinal items (Figure~\ref{fig:box_ord}) have a different
interpretation to those for the nominal items (Figure~\ref{fig:box_nom}).
The binary and ordinal responses have been coded with the convention
that larger responses correspond to greater wealth. Thus, a higher mean
value for the latent data corresponding to these items is indicative of
greater wealth. To interpret the box plots for the nominal items, all
latent dimensions for a particular item must be considered. If the mean
of one dimension ($k$, say) is greater than the means of the others
for
a particular cluster, then the response corresponding to dimension $k$
is the most likely response within that cluster. If the means for all
dimensions for a particular item are less than $0$, then the most
likely response by households in that cluster is the first
choice.

The box plots corresponding to the binary and ordinal items are shown
in Figure~\ref{fig:box_ord}. The elements of the mean of cluster 1 (the
wealthy/modern cluster) $\underline{\mu}{}_1$ can be seen to be
greater than those for the other clusters in general; this reflects the
greater wealth observed in cluster 1 compared to the other groups.
Similarly, the elements of $\underline{\mu}{}_3$ (the least wealthy
group) are lower than those for the other groups, reflecting the lower
socio-economic status of households in cluster 3. The difference
between cluster 3 and clusters 1 and 2 is particularly stark on the
location of toilet facilities (\emph{ToiletFac}) and the type of toilet
facilities (\emph{ToiletType}) items. The means for clusters 1 and 2
are notably higher than the mean for cluster 3 since the responses for
groups 1 and 2 are typically a number of categories higher on these items.

%

%

Figure~\ref{fig:box_nom} shows box plots of the MCMC samples of the
dimensions of the cluster mean parameters, $\underline{\mu}{}_g$,
corresponding to the nominal items. Focusing on
the latent dimensions corresponding to the \emph{PowerCook} item, say,
it can be seen that the highest mean for cluster 1 is on the
``electricity'' dimension followed closely by the ``wood'' dimension,
and that these means are greater than $0$. This implies that the most
likely response to the \emph{PowerCook} item for cluster 1 is
electricity but that a significant proportion of the households in this
group cook with wood. The highest means for clusters 2 and 3 are on the
``wood'' latent dimension. Thus, most of the households in these
clusters cook with wood in contrast to the wealthy/modern cluster 1.
This difference is indicative of a socio-economic divide. In a similar
way, the mean parameters for the \emph{PowerLight} item suggest that
electricity is the most likely source of power for lighting for
households in all clusters; the parameter estimates associated with the
\emph{Roof} item suggest corrugated iron roofs are the predominant
roofing type on dwellings in the Agincourt region.

To further investigate the difference between the $3$ clusters, the
response probabilities to individual survey items within a cluster are
examined. For example, Table~\ref{RespProbs} shows the probability of
observing each possible response to the \emph{Stove} item, conditional
on the members of each cluster.

\begin{table}
\tablewidth=150pt
\caption{Cluster specific response probabilities to
the survey item \emph{Stove}}
\label{RespProbs}
\begin{tabular*}{150pt}{@{\extracolsep{\fill}}lcc@{}}
\hline
\textit{\textbf{G}} & \textbf{No} & \textbf{Yes} \\
\hline
1 & 0.005 & 0.995 \\
2 & 0.509 & 0.491 \\
3 & 0.626 & 0.374 \\
\hline
\end{tabular*}
\end{table}

\begin{figure}

\includegraphics{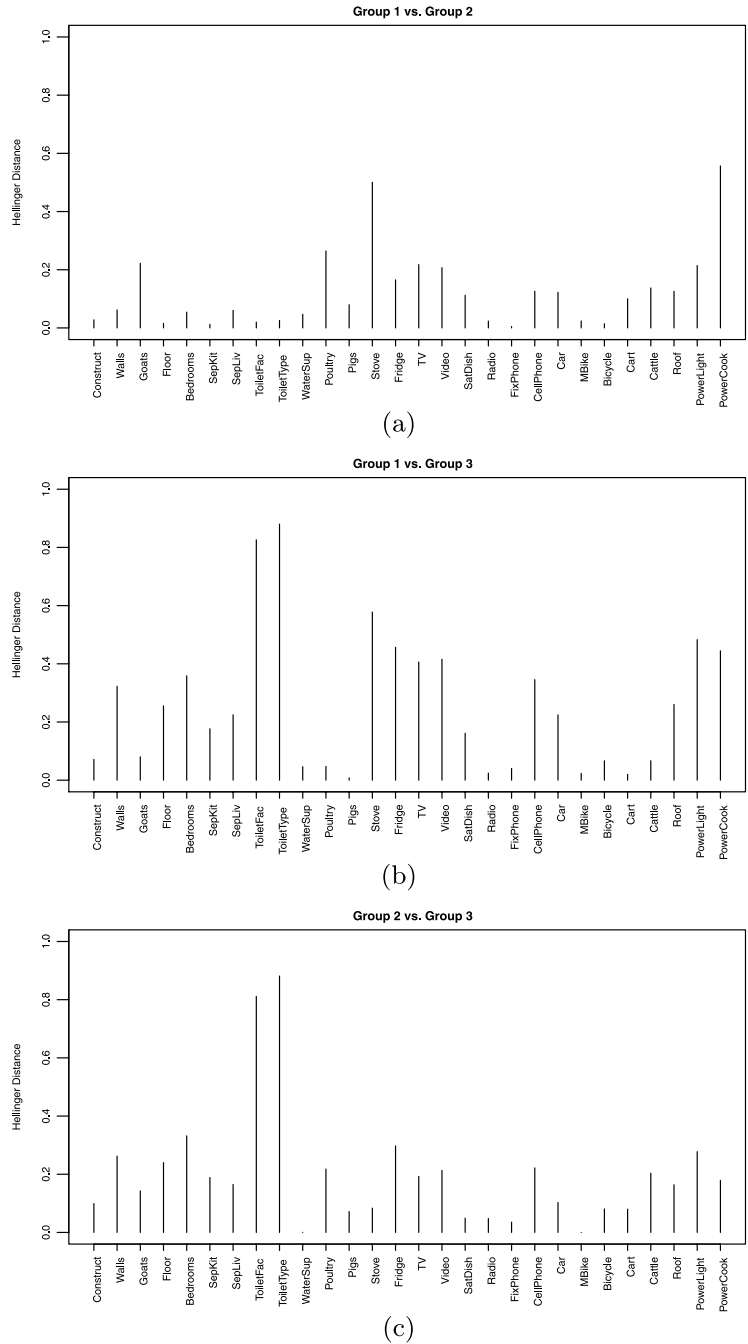}

\caption{Pairwise comparisons of groups using Hellinger distance.
\textup{(a)} Hellinger distances between groups 1 and 2. Total distance
is 3.544.
\textup{(b)} Hellinger distances between groups 1 and 3. Total distance is 7.316.
\textup{(c)} Hellinger distances between groups 2 and 3. Total distance
is 5.643.}
\label{fig:HelPlot}
\end{figure}

%
%

The distances between the cluster specific item response probability
vectors can be used to make pairwise comparisons of groups. The
distance measure
used here is Hellinger distance [\citet{lecam90,rao95,bishop06}].
Pairwise comparisons between clusters are illustrated in Figure~\ref{fig:HelPlot}. The Hellinger distance between response probability
vectors for each item is plotted. The groups that are most different
are clusters 1 (the wealthy/modern cluster) and 3 (the least wealthy
cluster). The sum of the Hellinger distances between these groups
across all items is 7.316. The items for which
the Hellinger distance between the response probability vectors is
largest are \emph{ToiletType}, \emph{ToiletFac}, \emph{Stove} and \emph
{PowerLight}, highlighting the areas in which households in these
clusters differ most. There are noteworthy
Hellinger distances for many other items also. The difference in
response patterns for these items is also evident in the box plots in
Figures~\ref{fig:box_ord} and~\ref{fig:box_nom}.

The sum of the Hellinger distances between clusters 1 and 2 (the
wealthier two clusters) across all items is 3.544, making these two
groups the most similar. There are some notable differences, however,
the Hellinger distance between the groups on the items \emph{Stove} and
\emph{PowerCook} are 0.501 and 0.556, respectively, which accounts for
almost $30\%$ of the total distance.

Clusters 2 and 3 are quite different and the sum of the Hellinger
distances between these groups is 5.643. As was the case for clusters 1
and 3, the
items \emph{ToiletType} and \emph{ToiletFac} provide the largest
Hellinger distances between groups 2 and~3. In contrast, however, there
are much smaller differences
for the items \emph{Stove} and \emph{PowerCook}. Again these results
highlight the specific areas in which the socio-economic status of
households within each cluster differ. A similar pattern was observed
in Table~\ref{modes3} and Figures~\ref{fig:box_ord} and~\ref{fig:box_nom}.

\begin{figure}[b]

\includegraphics{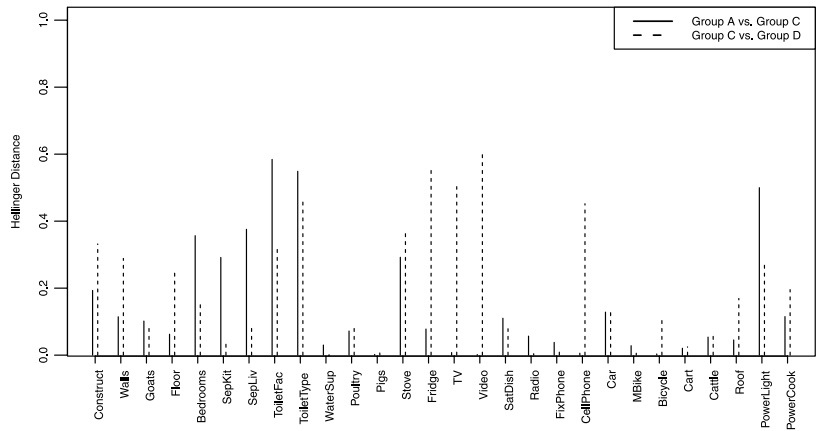}

\caption{Hellinger distances between groups A and C (red) and groups C
and D (blue). The total distances are 4.219 and 5.699 respectively.}
\label{fig:HelPlot4}
\end{figure}

\subsection{Results: Four- and two-component MFA-MD models}
\label{subsec:4CompRes}

Many of the substantive results returned by the 4-component model are
similar to those inferred from the 3-component model. Notably, the
items listed in Table~\ref{modes3} (i.e., those items for which the
modal response differs across groups in the 3-component solution) are a
subset of those items for which the modal response differs across
groups in the 4-component solution [details provided in \citet
{mcparland2014}]. Groups A, B and C in the 4-component model are
substantively similar to groups 1, 2 and 3 from the 3-component model,
respectively. Group D returned by the 4-component model is interesting,
however. It is similar to group A in that households in this cluster
possess many modern conveniences but the standard of their dwelling is
not at the same level as those in group A. The standard of dwellings in
group D is similar to those in group C, however, the households differ
from group C in terms of the modern conveniences they possess. Further
investigation revealed that households in group C are either
in group 1 (wealthy) or group 3 (poor) of the 3-component solution.
Figure~\ref{fig:HelPlot4} plots the Hellinger distance between groups A
and C and groups C and D, illustrating the differences and similarities
between these clusters. It can be seen that the largest distances
between groups A and C concern items related to the dwelling while the
largest distances between groups C and D concern modern convenience
ownership.\looseness=-1

Interestingly, group 2 and group B, under the 3- and 4-component
solutions, respectively, consist of almost exactly the same households.
These groups are deemed to be wealthy but less modern than group 1 and
group A, under the 3- and 4-component solutions, respectively. Indeed,
under the 5- and 6-component models, the essence of this cluster
remains intact.

Similar substantive results are inferred from the two-component MFA-MD
solution. Again, it is notable that those items for which the modal
response differs across groups in the 2-component solution [detailed in
\citet{mcparland2014}] are a subset of those items for which the modal
response differs across groups in the 3-component solution (detailed in
Table~\ref{modes3}). Groups A and B under the 2-component solution
relate generally to clusters 1 and 2 in the 3-component solution. The
poorer cluster B in the 2-component solution separates to create
clusters~2 and 3 in the 3-component solution.

\begin{figure}[b]\vspace*{-2pt}

\includegraphics{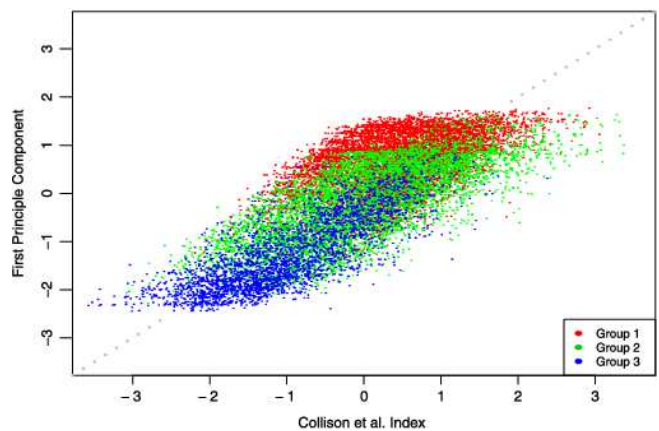}\vspace*{-2pt}

\caption{Comparing the principal component based approach of
Filmer and Pritchett (\citeyear{filmer01}) to the asset index of
Collinson et~al. (\citeyear{collinson09}). The gray line
shows where both scores are equal and the points are colored according
to the 3 group, 1-dimensional latent trait, MFA-MD solution.}
\label{fig:ComparePlot}
\end{figure}

\subsection{Comparison to existing methodology}
\label{subsec:compare}
Several other approaches to exploring the SES landscape based on asset
survey data are detailed in the demography literature. It is therefore
of interest to compare the results obtained when exploring the
Agincourt SES landscape using the proposed MFA-MD model to those
obtained when existing methods are applied. In particular, two existing
methods for analyzing mixed type socio-economic data are considered,
that of \citet{filmer01} and that of \citet{collinson09}, mentioned in
Section~\ref{sec:data}.

The \citet{filmer01} approach codes ordinal and nominal responses using
dummy binary variables and a principal component analysis (PCA) is
applied to the resulting data matrix. The \citet{collinson09}
approach
constructs a continuous asset index from the raw data. Figure~\ref{fig:ComparePlot} shows the standardized first principal component
scores plotted against the standardized asset index of \citet
{collinson09} when these methods were applied to the Agincourt data.
The points are colored by the three-group clustering solution
considered here. The two alternative scores do seem to broadly agree.
In addition, the 3-cluster solution appears to roughly correspond to
the gradation of the first principal component scores. However, \citet
{filmer01}\vadjust{\goodbreak} partition households into the lowest 40\%, middle 40\% and
top 20\% based on these principal component scores; their choice of
percentiles is arbitrary. Comparing the allocation based on this
criterion to that from our model results in a Rand and adjusted Rand
index of 0.61 and 0.15, respectively. Thus, the clustering solution using the
MFA-MD model is quite different than that currently in use. Clustering
households is not the primary goal for \citet{collinson09}, though they
do classify households as ``chronically poor'' if they have below the
median asset index score. The MFA-MD model allocates households using a
more preferable objective model-based approach, while recognizing the
different data types and treating them accordingly.

\section{Discussion}
\label{sec:discuss}

This paper set out to describe and understand the SES landscape in the
Agincourt region in South Africa through clustering households based on
their asset status survey data. The MFA-MD model described in this
paper successfully achieved this aim by clustering households into
groups of differing socio-economic status. Which households are in each
group and what differentiates these clusters from each other can be
examined in the model output. This information is potentially of great
benefit to various authorities in the Agincourt region. The
interpretation of the SES clusters could aid decision-making with
regard to infrastructural development
and other social policy. Further, the resulting clustering memberships
and cluster interpretations will be used to aid targeted sampling of a
particular cluster of households in the Agincourt region in future
surveys. New questions in future surveys can be derived based on the
substantive information now known about the SES clusters. The
clustering output from the MFA-MD model could also be used as covariate
input to other models, such as mortality models. There may be important
differences in mortality rates in different socio-economic strata
within the region; new health policies may need to take these
differences into account. A key interest for the sociologists studying
the Agincourt region is understanding social mobility, and
substantively examining SES clusters is the first step in this process.
Thus, the clustering exploration of the SES landscape in Agincourt will
provide support to researchers in the Agincourt region, through the
exposure of (dis)similar features of the clusters of households. The
information provided about the SES Agincourt landscape is based on a
statistically principled clustering approach, rather than ad hoc measures.

The MFA-MD model also provides a novel model-based approach to
clustering mixed categorical data. The SES data used here is a mix of binary,
ordinal and nominal data. The MFA-MD model provides clustering
capabilities in the context of such mixed data without mistreating any
one data type. A
factor analytic model is fitted to each group individually and may be
interpreted in the usual manner.

Future research directions are plentiful and varied. The lack of a
formal model selection criterion for the MFA-MD model is the
most\vadjust{\goodbreak}
pressing, and challenging. The provision of a formal criterion would
facilitate application of the MFA-MD model in settings in which an
optimal model must be selected; a formal criterion which selects the
most appropriate number of components and also the dimension of the
latent trait would be very beneficial. Model selection tools based on
the marginal likelihood [\citet{friel11}] are a natural approach to model
selection within the Bayesian paradigm, but the intractable likelihood
of the observed data $Y$ poses difficulties for the MFA-MD model. This
renders even approximate approaches such as BIC unusable. One
alternative would be to approximate the observed likelihood using the
underlying latent data $Z$, but this also brings difficulties and
uncertainty [\citet{mcparland13b}]. Other joint approaches to clustering
and choosing the number of components are popular in the
literature; using a Dirichlet process mixture model or incorporating
reversible jump MCMC may provide fruitful future research directions.
However, the applied nature of the work here and the requirement of
interpretative clusters and model parameters motivated the use of a
finite mixture model. Approaches to choosing the number of latent
factors such as those considered in \citet{lopes04} or \citet
{bhattacharya11} could also have potential within the MFA-MD context.

Additionally, there are several ways in which the MFA-MD model itself
could be extended. Here, the last time point from the Agincourt survey
was analyzed. However, there have been several waves of this particular
survey---extending the MFA-MD model to appropriately model
longitudinal data would be
beneficial. In this way the Agincourt households could be tracked
across time, as they may or may not move between socio-economic strata.
As with most clustering
models, the variables included in the model are potentially
influential. The addition of a variable selection method within the
context of the MFA-MD
model could significantly improve clustering performance and provide
substantive insight to asset indicators of SES. A~reduction in the
number of variables would also decrease the computational time required
to fit such models. In a similar vein, the Metropolis--Hastings step
required to sample the threshold parameters in the current model
fitting approach
could potentially be removed by using a rank likelihood approach [\citet
{hoff09}]. This
could also offer an improvement in computational time.\looseness=1

Other areas of ongoing and future work include the inclusion of
modeling continuous data by the MFA-MD model. This would facilitate the
clustering
of mixed data consisting of both continuous and categorical data [\citet
{mcparland13a}], and requires little extension to the MFA-MD model
proposed here. Allowing further correlations in the latent variable
beyond those produced by the latent trait is an interesting model
extension; this could be achieved by relaxing the unit variance in the
probit link.
Finally, covariate information could naturally be incorporated in the
MFA-MD model in the mixture of experts framework
[\citet{gormley08,jacobs91}]; such an approach could be insightful in understanding cause-effect
relationships in the Agincourt SES clusters and should be a
straightforward extension.

%
\begin{table}[b!]
\caption{A list of all survey items and the possible
responses. The final three items in the table are regarded as nominal,
all other items are binary or ordinal}
\label{tab:items}
{\fontsize{9}{10}\selectfont{
\begin{tabular*}{\textwidth}{@{\extracolsep{\fill}}lcc@{}}
\hline
\multicolumn{1}{@{}l}{\textbf{Item}} &
\textbf{Description} & \multicolumn{1}{c@{}}{\textbf{Response options}}\\
\hline
Construct & Indicates whether main dwelling is still & (No, Yes) \\
& under construction. &\\
Walls & Construction materials used for walls. & (Informal, Modern) \\
Floor & Construction materials used for floor. & (Informal, Modern)\\
Bedrooms& Number of bedrooms in the household. & ($\leq$1, 2, 3, 4, 5,
$\geq$6) \\
SepKit& Indicates whether kitchen is separate & (No, Yes) \\
& from sleeping area. & \\
SepLiv& Indicates whether living room is separate & (No, Yes) \\
&from sleeping area. &\\
ToiletFac& Reports the physical location of toilet in & (Bush, Other
House,\\
&the household.& In Yard, In House ) \\
ToiletType& Reports the type of toilet used. & (None, Pit, VIP,
Modern) \\
WaterSup& Reports the water supply source. & (From a tap, Other) \\
Stove& Reports stove ownership status. & (No, Yes) \\
Fridge& Reports fridge ownership status. & (No, Yes) \\
TV& Reports television ownership status. & (No, Yes) \\
Video& Reports video player ownership status. & (No, Yes) \\
SatDish& Reports satellite dish ownership status. & (No, Yes) \\
Radio& Reports radio ownership status. & (No, Yes) \\
FixPhone& Reports fixed phone ownership status. & (No, Yes) \\
CellPhone& Reports mobile phone ownership status. & (No, Yes) \\
Car& Reports car ownership status. & (No, Yes) \\
MBike& Reports motor bike ownership status. & (No, Yes) \\
Bicycle& Reports bicycle ownership status. & (No, Yes) \\
Cart& Reports animal drawn cart ownership & (No, Yes) \\
& status. & \\
Cattle& Reports cattle ownership status. & (No, Yes) \\
Goats& Reports goats ownership status. & (No, Yes) \\
Poultry& Reports poultry ownership status. & (No, Yes) \\
Pigs& Reports pig ownership status. & (No, Yes) \\
Roof& Construction materials used for roof. & (Other informal,
Thatch,\\
&& Other modern,\\
&& Corrugated iron, Tile)\\
PowerLight& Main power supply for lights & (Other, Candles, Paraffin,\\
& and appliances.& Solar, Battery/Generator,\\
&& Electricity)\\
PowerCook& Main power supply for cooking. & (Other, Wood, Paraffin,\\
&& Gas Bottle, Electricity) \\
\hline
\end{tabular*}}}
\end{table}

\begin{appendix}
\label{app}
\section{Survey items}
\label{app:items}

\newpage
\section{Latent variable formulation of nominal responses}
\label{app:toy}

Suppose item $j$ is nominal with $K_j=3$ options: apple (denoted level
1), banana (denoted level 2) or pear (denoted level 3). Thus,
$\underline{z}_{ij} = \{z_{ij}^1, z_{ij}^2\}$.\vspace*{2pt}


%
%
%
%
%
%

\begin{figure}

\includegraphics{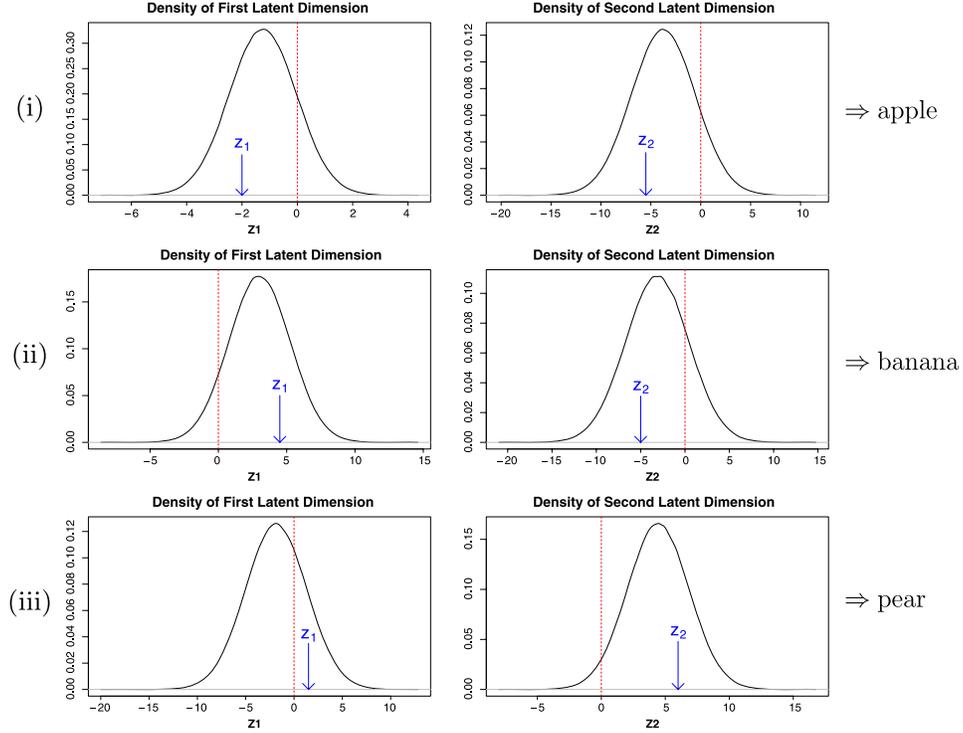}

\caption{Latent variable formulation of nominal responses.}
\label{fig:ToyPlot}\vspace*{12pt}
\end{figure}

Figure~\ref{fig:ToyPlot} shows what the marginal distributions of the
latent variables might look like along with realizations from those
distributions:
\begin{longlist}[(iii)]
\item[(i)] Both $z_{ij}^1$ and $z_{ij}^2$ are less than 0, thus
 $\max_k\{z_{ij}^k\} < 0 \Rightarrow y_{ij} = 1$, that is, apple.\vspace*{1pt}
\item[(ii)] $z_{ij}^1 = \max_k\{z_{ij}^k\}$ and $z_{ij}^1 > 0
\Rightarrow y_{ij} = 2, \mbox{ that is, banana}$.\vspace*{1pt}
\item[(iii)] $z_{ij}^2 = \max_k\{z_{ij}^k\}$ and $z_{ij}^2 > 0
\Rightarrow y_{ij} = 3, \mbox{ that is, pear}$.\vspace*{1pt}
\end{longlist}
In the MCMC algorithm, these latent variables are sampled conditional
on the observed data $Y$. Given the nominal response, the full
conditional distributions are truncated appropriately.\newpage

\section{Convergence of Markov chains}
\label{app:cvg}

\end{appendix}

\begin{figure}[b!]

\includegraphics{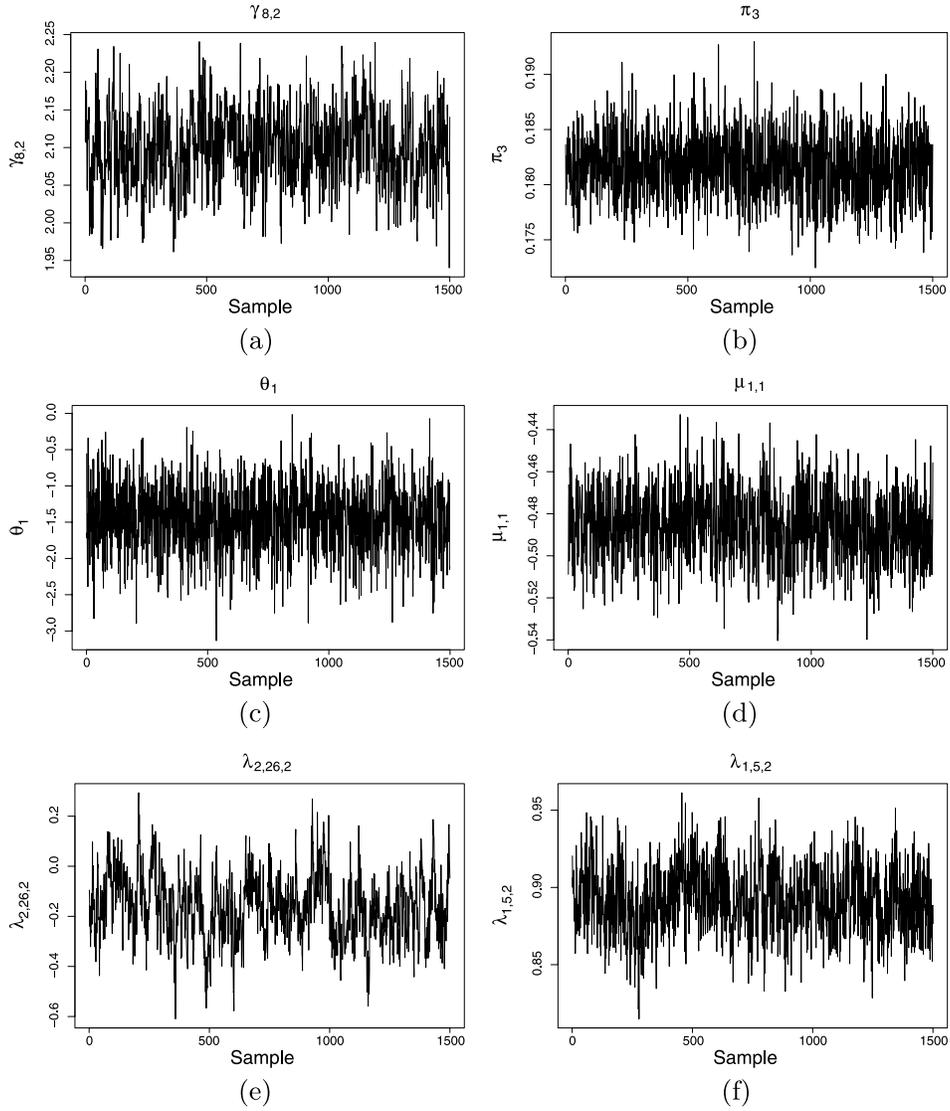}

\caption{Trace plots of Markov chains for selected parameters. The
plots shown are of the thinned MCMC samples, post burn-in.
\textup{(a)} Trace plot of the MCMC samples for one of the threshold
parameters of the $\mathit{ToiletFac}$ item.
\textup{(b)} Trace plot of the MCMC samples of the mixing weight for
group 3.
\textup{(c)} Trace plot of the MCMC samples for the latent trait of the
first household.
\textup{(d)} Trace plot of the MCMC samples of the first dimension of
the mean vector for group 1.
\textup{(e)} Trace plot of the MCMC samples of one of the loadings
parameters for nominal item $\mathit{Roof}$.
\textup{(f)} Trace plot of the MCMC samples of the loadings parameter
for ordinal item $\mathit{Bedrooms}$.}\label{fig:cvg}
\end{figure}


\section*{Acknowledgments}

The authors wish to thank Professor Brendan Murphy, Professor Adrian
Raftery, the members of the Working Group on Statistical Learning at
University College Dublin and the members of the
Working Group on Model-based Clustering at the University of Washington
for numerous suggestions that contributed enormously to this work.

\begin{supplement}[id=suppA]
\sname{Supplement A}
\stitle{Full conditional posterior distributions\\}
\slink[doi]{10.1214/14-AOAS726SUPPA} 
\sdatatype{.pdf}
\sfilename{aoas726\_suppA.pdf}
\sdescription{Derivations of the full conditional posterior distributions.}
\end{supplement}

\begin{supplement}[id=suppB]
\sname{Supplement B}
\stitle{Additional results\\}
\slink[doi]{10.1214/14-AOAS726SUPPB} 
\sdatatype{.pdf}
\sfilename{aoas726\_suppB.pdf}
\sdescription{Additional tSSPR sensitivity analysis, Bayesian latent
residual plots and tables of results.}
\end{supplement}

\begin{supplement}[id=suppC]
\sname{Supplement C}
\stitle{C code\\}
\slink[doi]{10.1214/14-AOAS726SUPPC} 
\sdatatype{.zip}
\sfilename{aoas726\_suppC.zip}
\sdescription{C code to fit the MFA-MD model for clustering in the
context of mixed categorical data.}
\end{supplement}



\vspace*{25pt}

\printaddresses

\end{document}